\documentclass[conference]{IEEEtran}
\IEEEoverridecommandlockouts
\usepackage{cite}
\usepackage{amsmath,amssymb,amsfonts}
\usepackage{algorithmic}
\usepackage{graphicx}
\usepackage{textcomp}
\usepackage{xcolor}
\def\BibTeX{{\rm B\kern-.05em{\sc i\kern-.025em b}\kern-.08em
    T\kern-.1667em\lower.7ex\hbox{E}\kern-.125emX}}
\begin{document}

\title{Learnable MFCCs for Speaker Verification\\

\thanks{This work was partially supported by Academy of Finland (project 309629) and Inria Nancy Grand Est.}
}

\author{\IEEEauthorblockN{Xuechen Liu}
\IEEEauthorblockA{\textit{CNRS, Inria, LORIA} \\
\textit{Universit\'{e} de Lorraine}\\
F-54000, Nancy, France \\
xuechen.liu@inria.fr}
\and
\IEEEauthorblockN{Md Sahidullah}
\IEEEauthorblockA{\textit{CNRS, Inria, LORIA} \\
\textit{Universit\'{e} de Lorraine}\\
F-54000, Nancy, France \\
md.sahidullah@inria.fr}
\and
\IEEEauthorblockN{Tomi Kinnunen}
\IEEEauthorblockA{\textit{School of Computing} \\
\textit{University of Eastern Finland}\\
FI-80101 Joensuu, Finland \\
tkinnu@cs.uef.fi}
}

\maketitle

\begin{abstract}
We propose a learnable mel-frequency cepstral coefficients (MFCCs) front-end architecture for deep neural network (DNN) based automatic speaker verification. Our architecture retains the simplicity and interpretability of MFCC-based features while allowing the model to be adapted to data flexibly. In practice, we formulate data-driven version of four linear transforms in a standard MFCC extractor --- windowing, discrete Fourier transform (DFT), mel filterbank and discrete cosine transform (DCT). Results reported reach up to 6.7\% (VoxCeleb1) and 9.7\% (SITW) relative improvement in term of equal error rate (EER) from static MFCCs, without additional tuning effort.
\end{abstract}


\section{Introduction}
\emph{Automatic speaker verification} (ASV) \cite{Hansen_summaryasv2015} is used in forensic voice comparison, personalization of voice-based services and, more recently, smart home electronic devices. A typical ASV system can be broken down into three elementary components: (i) a frame-level feature extractor, (ii) a speaker embedding extractor, and (iii) a speaker comparator. Their functions are, respectively, to transform a waveform into a sequence of feature vectors, to extract fixed-sized speaker embedding vectors, and to compare two speaker embeddings (one from an enrollment and the other one from a test utterance).

While previous generations of ASV technology relied largely on statistical approaches such as i-vectors
\cite{Dehak_ivector2011}, state-of-the-art ASV leverages from
\emph{deep neural networks} (DNNs) to extract speaker embeddings.
Representative examples include \emph{d-vector} \cite{Variani_dvector2014} and \emph{x-vector} \cite{Snyder_xvector2018}. Numbers of extensions from them have been proposed as well \cite{You_multitask_xvector2019, Snyder_extended_xvector2019}. Common to most of those speaker embedding extractors is using either \emph{mel-frequency cepstral coefficients} (MFCCs) \cite{Davis80-MFCC} or raw spectrum as frame-level features. Different from the speaker embedding extractor, whose parameters are obtained through numerical optimization, raw spectra and MFCCs are obtained with fixed/static operations. In this work, our main goal is to formulate a lightweight, data-driven version of a standard MFCC extractor.

Related recent work includes so-called \emph{end-to-end} \cite{Ravanelli_sincnet2018}\cite{Zeghidour_learning_fb_2018} front-end solutions. Using DNN-based components that are optimized jointly, such end-to-end solutions process the raw waveform to produce either detection scores or intermediate features to be used with other components. Despite promising results, the end-to-end approaches tend to require substantial engineering efforts, making them potentially inflexible for adaptation to new applications or data. Additionally, unless some prior domain knowledge is used in designing the DNN components, such models can be difficult to interpret. Meanwhile, analysis and assessment of relative importance of different signal processing components is important in speech-related research. Interpretability is also demanded in high-stake applications, such as forensic voice comparison. 

For reasons above, we advocate a novel architecture whose design is guided by one the most successful fixed feature extractor, MFCCs. Even if an MFCC extractor is typically not viewed as a neural network, it can be seen as a DNN consisting of a number of linear layers (and some non-linearities). It is therefore a natural idea to expand the speaker embedding extractor to include MFCC-specific layers to be optimized in a data-driven manner. This is enabled by defining a computational graph and the associated automatic differentiation procedures available in standard DNN toolboxes. Though we draw  inspiration from similar ideas in other tasks (e.g. \cite{vasquez_melnet_2019, Pariente_fbdesign_icassp2020}), our aim is an initial formulation and an experimental feasibility study in the context of ASV.

\section{learnable MFCC extractor}
\label{sec:methodology}

\subsection{Front-end MFCC extractor}
\label{speech_frontend}
A typical MFCC extractor consists of a cascade of linear and non-linear transformations originally motivated \cite{Davis80-MFCC} from signal processing and human auditory system considerations. Typical steps (after pre-processing) include windowing, power spectrum computation using \emph{discrete Fourier transform} (DFT), smoothing by a bank of triangular-shaped filters, logarithmic compression and \emph{discrete cosine transform} (DCT). 


MFCCs have been used across many different speech and audio applications successfully, suggesting their generality as an application-agnostic frame-level feature. Nonetheless, the standard transformations in MFCC extractors may be improved further. For instance, \cite{Tomi_multitaper2012} uses low-variance multi-taper spectrum estimation to replace Hamming-windowed DFT. Other studies employ alternative time-frequency representations, such as \emph{constant-Q transform} (CQT)~\cite{Todisco+2016} and wavelet-based methods~\cite{Anden_deepscattring_2014,farooq2001mel}. Different frequency warping scales are studied in ~\cite{Umesh_melscale_1999}. Similarly, triangular filterbank can be replaced with Gaussian and Gammatone filterbanks~\cite{Kim_pncc2016}. The logarithmic compression can also be substituted with cube-root compression \cite{Hermansky_plp1990}. The suitability of block DCT as an alternative of standard DCT (i.e, DCT-II) is explored in \cite{Naveena_blockdct_2017}. 

All above studies focus on developing other fixed operation models by overcoming some of the limitations of the existing one. Differently from those studies, we propose to optimize the parameters of MFCC pipeline in a data-driven manner. We consider making learnable components based on static MFCCs only, as dynamic (delta) coefficients were not found useful in our previous work \cite{Xuechen_features2020}.



\subsection{Differentiable linear transforms for MFCCs}
\label{sec:diff_trans}
With the above motivations, we would like to start from fixed MFCCs by making four highlighted differentiable linear transforms learnable. Three of them are real-valued, namely, windowing, mel filterbank and DCT. Therefore, when designing their learnable counterparts, for each component we simply create operators that have the same input and output as the static one so that we retain the same exact computational flow. The only difference from static feature extractor is that the gradient can now be back-propagated to update the numerical values of the linear transforms. DFT is an exception since it is a \emph{complex}-valued linear operator. Nonetheless, when integrated as a step to produce a power spectrum, the operation can be expressed as:

\begin{equation}
    \begin{gathered}
    |\boldsymbol{X}|^2 = |\boldsymbol{Fx}|^{2} = |(\boldsymbol{F}_\text{real} + j\boldsymbol{F}_\text{imag})\boldsymbol{x}|^{2}\\ 
    = g(\boldsymbol{F}_\text{real}\boldsymbol{x}) + g(\boldsymbol{F}_\text{imag}\boldsymbol{x}),
    \end{gathered}
\label{eq:dft}
\end{equation}

\noindent where $g(\boldsymbol{Fx}) = |\boldsymbol{Fx}|^2$. Here, $\boldsymbol{x}$ is a windowed speech frame, $\boldsymbol{X}$ is the complex-valued spectrum, $\boldsymbol{F}$ is the complex-valued DFT matrix and $|.|^2$ denotes element-wise modulus. Thus, \eqref{eq:dft} can be implemented as two real-valued linear transforms, followed by squared summation.

\section{Optimization of learnable components}
We describe below three techniques to optimize the selected components. We refer to the corresponding matrices as \emph{kernels}, denoted by specific symbols: $\boldsymbol{W}$ for the window function, $\boldsymbol{F}$ for DFT (as noted in Equation \ref{eq:dft}), $\boldsymbol{M}$ for the mel filterbank, and $\boldsymbol{D}$ for DCT.

\subsection{Kernelized initialization}
Trainable component are often initialized using random numbers from a normal distribution \cite{xavier_normal_glorot2010}. In this work, however, we assert that a standard MFCC extractor serves as a reasonable starting point for further learning. Thus, our first technique initializes each kernel with its corresponding static counterpart. For windowing, we use the Hamming window \cite{Harris_windowing1978}. For mel filterbank and DCT static correspondents are available and can be directly used in place. For DFT, we generate kernels from the DFT matrix and separate the real and imaginary parts $\boldsymbol{F}_\mathrm{real}$ and $\boldsymbol{F}_\mathrm{imag}$ in Eq. \eqref{eq:dft}. After initialization, training proceeds the same way as for any standard DNN-based speaker embedding extractor. 

The kernel initialization sets a starting point for further adaptation. We consider two additions to the training procedure. The idea in both is to promote specific numerical properties of each static component to regulate learning, discouraging overly aggressive deviation from their respective static counterparts. We detail the two ideas, loss regularization and kernel update, in the following two subsections.

\subsection{Loss regularization} 
We modify the training objective of the speaker embedding extractor as $\mathcal{L}_{\mathrm{new}} = \mathcal{L} + \lambda * g_{\mathrm{loss}}(\boldsymbol{K})$,  where $\mathcal{L}$ is multi-class cross-entropy loss, $\mathcal{L}_\mathrm{new}$ is regularized loss, $\boldsymbol{K}$ denotes the kernel, and $\lambda$ is regularization constant. For all experiments addressed in this work, we set $\lambda = 0.1$. In Section \ref{sec:results}, systems adapted with such method are marked as \emph{name + loss.}, where \emph{name} is the name of adapted component. We design separate regularizer $g_\mathrm{loss}(\cdot)$ for each of the four linear components.

\textbf{Windowing}. Many window functions (e.g. Hamming and Blackman) are generated using sinusoids \cite{Harris_windowing1978}. Thus, our regularizer measures distance from the learnable window to a cosine function: $g_{\mathrm{loss}}(\boldsymbol{W}) = ||\boldsymbol{W}_{\mathrm{norm}} - \boldsymbol{C}||$, where $\boldsymbol{C}(n)=-\cos(2\pi n/M), \; n\in[0, M-1]$ is a cosine function, $M$ being equal to frame length (i.e. length of window vector), $\boldsymbol{W}_{\mathrm{norm}}$ is a mean-normalized window, and $||.||$ denotes Frobenius norm \cite{matrix_comp_1996}. 
Therefore, when the constraint equals zero, the window equals a cosine function.
    
\textbf{DFT}. A DFT matrix is squared and symmetric. It can be split into real and imaginary parts, both of which are real-valued, squared and symmetric. We therefore introduce such property when implementing regularization by computing matrix-wise distance of the kernel to its symmetric version: $\boldsymbol{F}_\mathrm{dist.} = \boldsymbol{F}_\mathrm{norm} - \boldsymbol{F}_\mathrm{norm}\boldsymbol{F}_\mathrm{norm}^\top$, where $\boldsymbol{F}_\mathrm{dist.}$ is the difference matrix and $\boldsymbol{F}_\mathrm{norm}$ is the normalized version of $\boldsymbol{F}$. This applies to both $\boldsymbol{F}_\mathrm{real}$ and $\boldsymbol{F}_\mathrm{imag}$ in Eq. \ref{eq:dft}. The Frobenius norm of $\boldsymbol{F}_\text{dist.}$ is then used for regularization: $g_\mathrm{loss}(\boldsymbol{F}) = ||\boldsymbol{F}_\mathrm{dist.}||$. Therefore, when the constraint is perfectly met ($g_\mathrm{loss}(\boldsymbol{F}) = 0$), we see that $\boldsymbol{F}_\mathrm{norm}^\top = \boldsymbol{I}$, where $\boldsymbol{I}$ is an identity matrix.
    
\textbf{Mel filterbank}. Mel filterbank is a set of overlapped triangular filters with scaled peak magnitude, which can be either constant across all filters (our case) or varied via different frequency bins \cite{Rabiner_asrtextbook_1993}. Computationally, it is a matrix with non-negative elements with high sparsity. In order to control the level of sparsity of the kernel, we adopt $L_2$ regularization \cite{hastie_esl_2001} on the filterbank kernel to avoid over-fitting, instead of $L_1$, which tends to have a more enhancing effect on sparsity of model as a loss regularizer. Formally, $g_\mathrm{loss}(\boldsymbol{M}) = ||\boldsymbol{M}||^2$.
    
\textbf{DCT}. A DCT matrix is orthonormal, i.e. $\boldsymbol{D}\boldsymbol{D}^\top = \boldsymbol{D}^\top\boldsymbol{D} = \boldsymbol{I}$. 
We employed a recently-proposed soft orthonormality loss function \cite{ortho_dctloss_zhu2020}, expressed as $g_\mathrm{loss}(\boldsymbol{D}) = \Vert\boldsymbol{D}^\top\boldsymbol{D} - \boldsymbol{I}\Vert^2$, where 
$\boldsymbol{I}$ is the identity matrix. Optimizing such loss function minimizes the distance between the Gram matrix of $\boldsymbol{D}$ and $\boldsymbol{I}$ to encourage orthonormality. 

\subsection{Kernel update}
Aside from loss regularization, the other optimization technique performs direct update on the kernel operators every time after gradient update. Compared to loss regularization, it is a more `brute-force' approach. The updated kernel matrix or vector is then directly used for next iteration: $ \boldsymbol{K}_\mathrm{new} = g_\mathrm{kernel}(\boldsymbol{K})$,
where, $\boldsymbol{K}$ is kernel matrix after gradient update and $\boldsymbol{K}_\mathrm{new}$ is the directly-updated one used for next iteration. In Section \ref{sec:results}, systems adapted with this method are marked as \emph{name + kernel.}, where \emph{name} is the name of adapted component. Design of updater $g_{\mathrm{kernel}}(.)$ for each component is as follows.

\textbf{Windowing}. Commonly-used window functions are non-negative and symmetric. Inspired by such properties, our kernel update is 
$g_\mathrm{kernel}(\boldsymbol{W}) = |\mathrm{cat}(\boldsymbol{W}_\mathrm{[:size/2]}, \boldsymbol{W}_{\mathrm{flip}})|$, where $\boldsymbol{W}_\mathrm{[:size/2]}$ denotes the half-size truncated version of window vector $\boldsymbol{W}$ while $\boldsymbol{W}_\mathrm{flip}$ is its flipped (time-reversed) version. Here $\mathrm{cat.}$ performs column-wise concatenation and $|.|$ denotes absolute values.
    
\textbf{DFT}. As noted above, DFT matrices are square and symmetric. To enhance such properties, we perform a simple update on the kernel: $g_\mathrm{kernel}(\boldsymbol{F}) = \boldsymbol{F}\boldsymbol{F}^\top$, where $\boldsymbol{F}$ and $\boldsymbol{F}_\mathrm{new}$ denote kernel at the end of the current iteration and the next iteration, respectively. It is easy to see that $\boldsymbol{F}_\mathrm{new}$ is indeed symmetric\footnote{$(\boldsymbol{F}\boldsymbol{F}^\top)^\top = (\boldsymbol{F}^\top)^\top\boldsymbol{F}^\top = \boldsymbol{F}\boldsymbol{F}^\top$.}. Similar to the loss regularization scheme, this update is applied to both $\boldsymbol{F}_\mathrm{real}$ and $\boldsymbol{F}_\mathrm{imag}$.
    
\textbf{Mel filterbank}. As mel filterbank is a set of overlapped triangular filters with non-negative values, we force positivity by replacing negative elements with a small value: $\forall i,j, \; g_\mathrm{kernel}(\boldsymbol{M}_{i,j}) = \epsilon \; \mathrm{if} \; \boldsymbol{M}_{i,j} \leq 0, \; \mathrm{otherwise} \; \boldsymbol{M}_{i,j}$, where $\epsilon = 10^{-4}$, $i$ and $j$ denote row and column indices of the filterbank $\boldsymbol{M}$.

\textbf{DCT}. For DCT, we again capture its orthonormality requirement from its static correspondent by performing QR decomposition~\cite{matrix_comp_1996} on the learnt kernel matrix: $g_\mathrm{kernel}(\boldsymbol{D}) = QR(\boldsymbol{D})$, where $QR(.)$ decomposes $\boldsymbol{D} = \boldsymbol{QR}$ and outputs only the orthogonal matrix $\boldsymbol{Q}$.
Such an operation can be performed because 
the kernel learnt corresponds to DCT is set to be a square matrix, which means number of mel filters is same as number of output cepstral coefficients. We acquire such design choice because setting number of filters same as final static feature dimension can bring competitive performance, as shown in \cite{Xuechen_features2020}. This applies to all experiments in this work, including baseline.

\section{Data and experimental protocol}
\label{sec:experiment}

\subsection{Data}
\label{sec:data}
We trained baseline x-vector model on \emph{dev} \cite{Nagrani_vox1_2017} partition of VoxCeleb1, which consists of 1211 speakers. We used the same dataset for additional training steps on learnable linear components. For evaluation, we considered one matched and another relatively mismatched condition. For the former, we used the \emph{test} partition of VoxCeleb1 that consists of 40 speakers, 18860 genuine trials and same number of impostor trials \cite{Nagrani_vox1_2017}. The latter was composed of the \emph{development} part of \emph{speakers-in-the-wild} \cite{Mclaren_sitw2016} (SITW) corpus ``core-core'' condition, containing 119 speakers. It contains 2597 genuine and 335629 impostor trials. We refer to the two datasets as \emph{Voxceleb1-test} and \emph{SITW-DEV}, respectively.

\subsection{System configuration}
\label{sec:baseline}
For the baseline system, we used 30 static MFCCs as the input features and replicated x-vector configuration from \cite{Snyder_xvector2018} as the speaker embedding extractor. We trained the model using VoxCeleb1 without any data augmentation and Adam \cite{Kingma_adam_2015} as optimizer. During test time, we extracted 512 dimensional speaker embedding from the first fully-connected layer after statistics pooling. 

We adapted each of the four learnable front-end systems at a time, using same data as for training. In order to prevent distractions in terms of joint optimization from scratch and meet the aim of providing light-weighted interface for adaptation, the selected component was jointly optimized with the pre-trained baseline x-vector.
Speaker embeddings for all systems with learnt front-end components were extracted in same manner as baseline after adaptation. 

For all systems, we applied energy-based \emph{speech activity detection} (SAD) before feature processor and \emph{cepstral mean normalization} (CMN). All embeddings extracted at inference time were length-normalized and centered prior to being transformed by a 200-dimensional \emph{linear discriminant analysis} (LDA). Scoring was implemented through \emph{probabilistic linear discriminant analysis} (PLDA) \cite{Ioffe_plda2006} classifier. We used Kaldi for data preparation and PLDA training and PyTorch \cite{Paszke_pytorch_2017} for all DNN-related training and inference experiments.

\subsection{Evaluation}
\label{sec:eval}
\emph{Equal error rate} (EER) and \emph{minimum detection cost function} (minDCF) were used to measure ASV performance. MinDCF was computed with target speaker prior $p=0.001$ and detection costs $C_\text{FA}=C_\text{miss}=1.0$. We used BOSARIS \cite{bosaris} to produce selected detection trade-off (DET) curves.

\begin{figure}[h]
    \centering
    \includegraphics[scale=0.4]{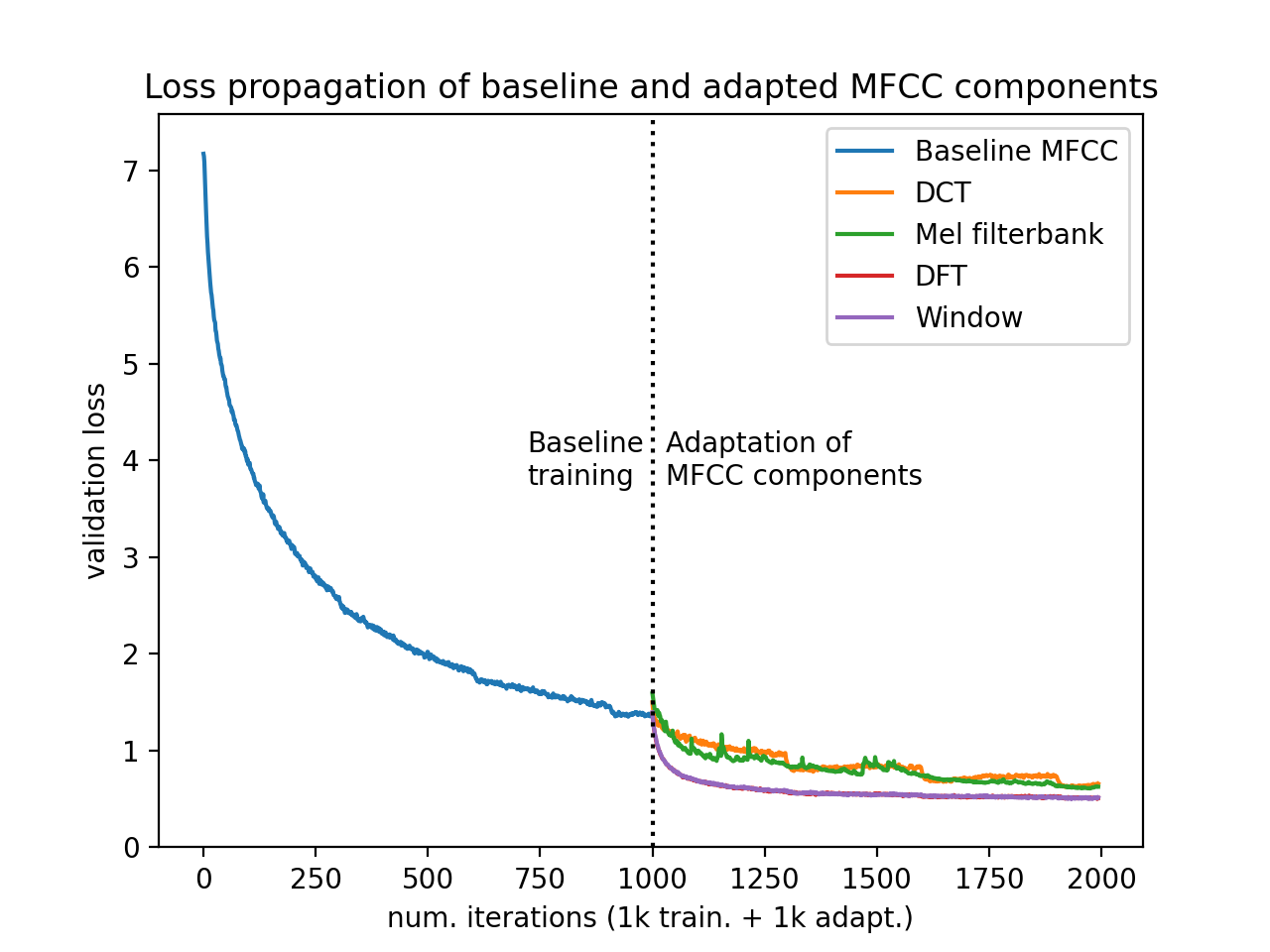}
    \caption{Loss propagation for baseline and adapted MFCC components. Best viewed in color.}
    \label{fig:loss_prop}
\vspace{-0.5cm}
\end{figure}

\begin{table}[htbp]
    \centering
    \caption{Results on \emph{Voxceleb1-test} and \emph{SITW-DEV}. All systems aside from baseline are with kernel initialization.}
    \vspace{-0.2cm}
    {\footnotesize %
    \begin{tabular}{|c|c|c|c|c|}
    \hline
         & \multicolumn{2}{c|}{Voxceleb1-test} & \multicolumn{2}{c|}{SITW-DEV} \\ \hline
         Operator (+optimal.) & EER(\%) & minDCF & EER(\%) & minDCF \\ \hline 
         Baseline MFCC & 4.64 & 0.6071 & 6.72 & 0.8243 \\ \hline 
         Window & 4.51 & 0.5544 & \textbf{6.09} & 0.7698 \\ 
         Window + loss. & 4.40 & 0.5508 & 6.66 & 0.7915 \\ 
         Window + kernel. & 4.42 & 0.5459 & \textbf{6.09} & 0.7819 \\ \hline
         DFT & \textbf{4.33} & 0.5933 & 6.35 & 0.7698 \\ 
         DFT + loss. & 4.71 & 0.6156 & 6.62 & 0.8041 \\ 
         DFT + kernel. & 5.42 & 0.6182 & 7.04 & 0.7903 \\ \hline 
         Melbank & 4.83 & 0.5767 & 6.52 & 0.8253 \\ 
         Melbank + loss. & 4.63 & 0.6162 & 6.39 & 0.8011 \\ 
         Melbank + kernel. & 4.45 & 0.5768 & 6.31 & \textbf{0.7689} \\ \hline 
         DCT & 4.36 & 0.5572 & 6.27 & 0.7950 \\ 
         DCT + loss. & 4.46 & \textbf{0.4971} & 6.54 & 0.7982 \\ 
         DCT + kernel. & 4.41 & 0.5501 & 6.54 & 0.7925 \\ \hline 
    \end{tabular}}%
    \label{tab:results}
\end{table}{}

\begin{figure*}[h]

\begin{minipage}[b]{0.2\linewidth}
  \centering
  \centerline{\includegraphics[width=5.5cm]{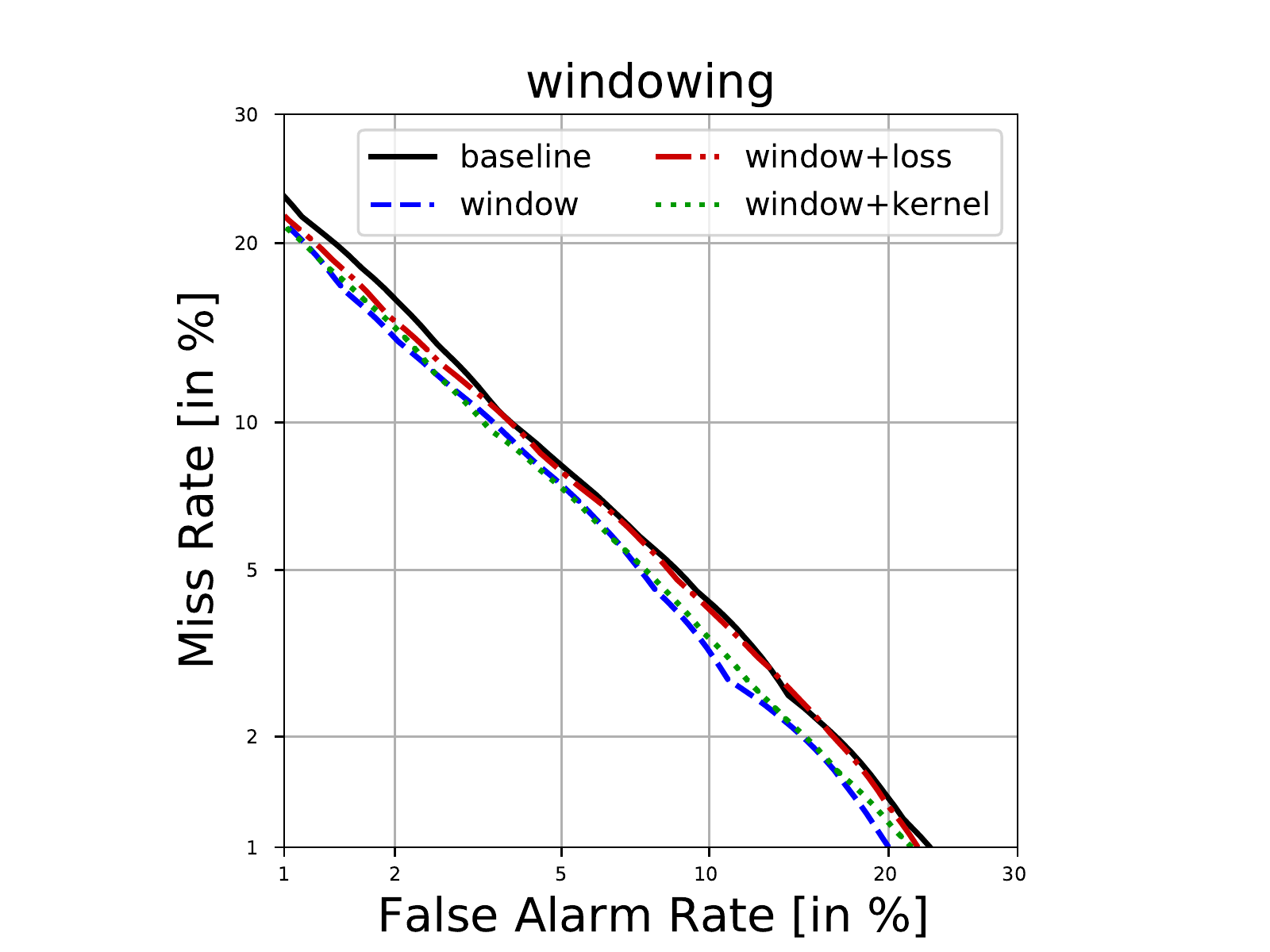}}
  \centerline{(a) windowing}\medskip
\end{minipage}
\hfill
\begin{minipage}[b]{0.2\linewidth}
  \centering
  \centerline{\includegraphics[width=5.5cm]{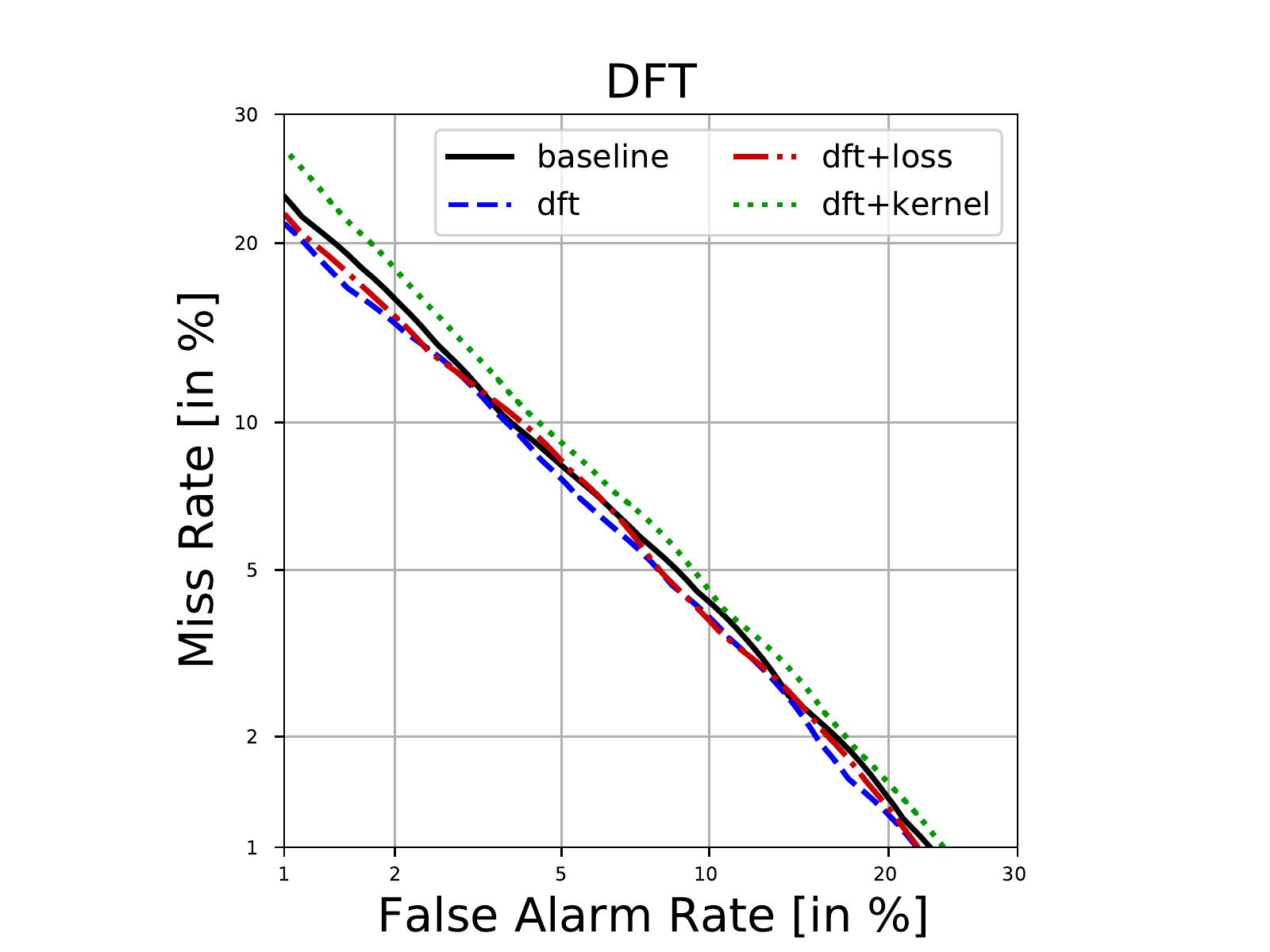}}
  \centerline{(b) DFT}\medskip
\end{minipage}
\hfill
\begin{minipage}[b]{0.2\linewidth}
  \centering
  \centerline{\includegraphics[width=5.5cm]{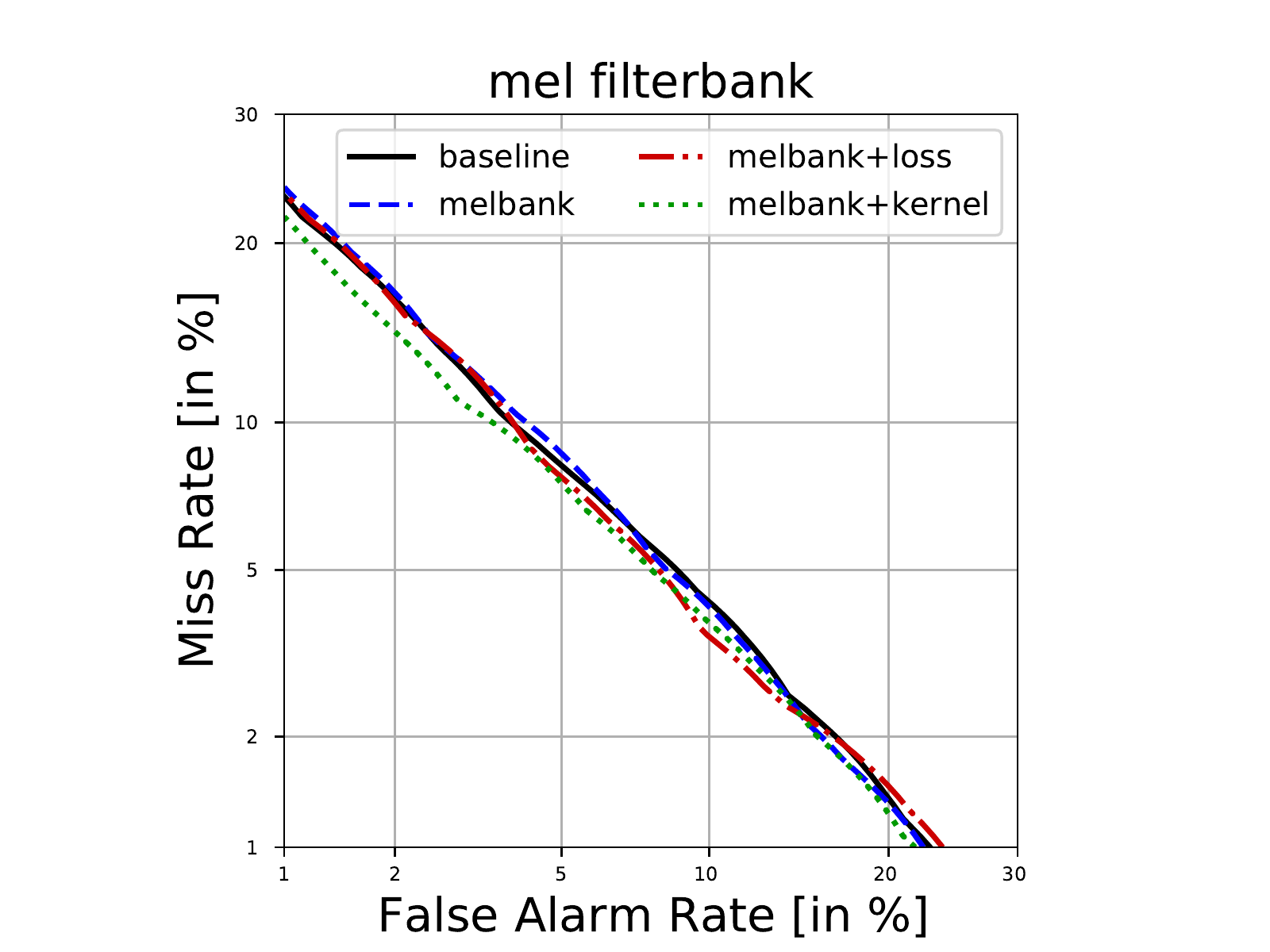}}
  \centerline{(c) mel filterbank}\medskip
\end{minipage}
\hfill
\begin{minipage}[b]{0.2\linewidth}
  \centering
  \centerline{\includegraphics[width=5.5cm]{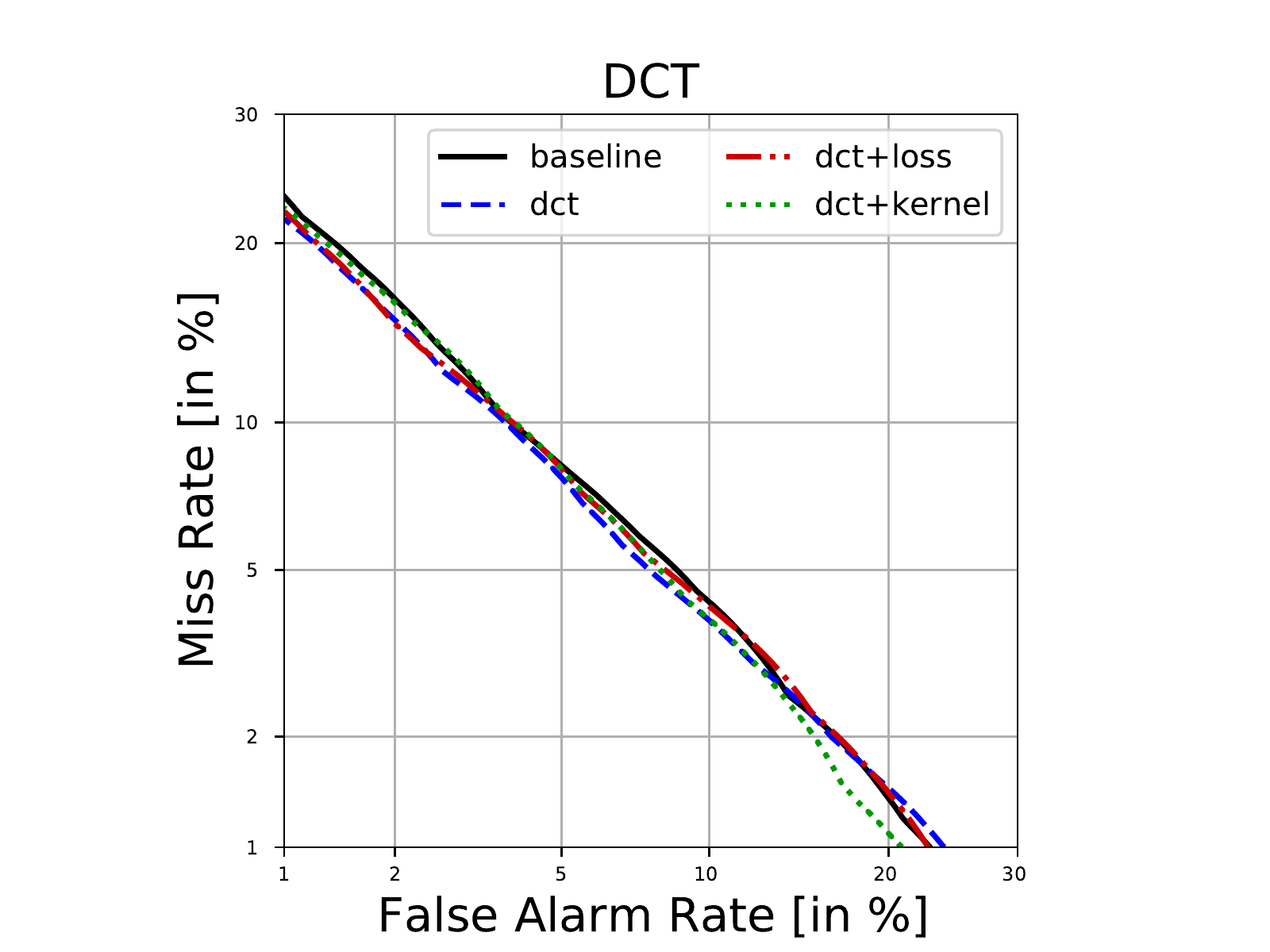}}
  \centerline{(d) DCT}\medskip
\end{minipage}
\hfill
\caption{DET plots for \emph{SITW-DEV}. Best viewed in color.}
\label{fig:det}
\vspace{-0.5cm}
\end{figure*}

\section{Results}
\label{sec:results}

Before presenting ASV results, we demonstrate validation loss (on \emph{dev} set) propagation of our baseline and adapted systems
in Fig.~\ref{fig:loss_prop}. The baseline x-vector system (with fixed MFCC components) was pre-trained with 1000 iterations, followed up by another 1000 iterations to adapt the MFCC components. 
The adaptation, especially for window function and DFT, results in a notable decrease of validation loss. This indicates potential to make components of an MFCC extractor learnable. The ASV results are reported in Table \ref{tab:results}. 

\subsection{ASV results on Voxceleb1-test}
Concerning windowing, all the three adapted variants outperform the baseline. Loss regularization and kernel update are particularly more effective. The results indicate usefulness to retain symmetricity and positivity of the window. 

Concerning DFT, simply letting it to be data-driven (without added regularization or kernel update) yields lowest EER among all systems, with a relative improvement of 6.7\% over the baseline.
In fact, the additional regularization and direct update are detrimental. This indicates potential weakness of our symmetricity constraint.

Concerning the mel filterbank, 
\emph{Melbank + kernel.} yields the best performance among the three adapted variants, with best minDCF of all systems, improving baseline by relatively 6.7\%. This indicates the importance of enforcing positivity of the learnt filters. 

Concerning DCT, similar to windowing all the learning schemes improve upon the baseline. While QR decomposition does not bring notable positive impact, the orthonormality-enhancing loss regularization results in slightly worse EER, but improved minDCF. In fact, \emph{DCT + loss.} results in lowest minDCF among all systems. 


\subsection{ASV results on SITW-DEV}

We now move on to discuss ASV results on the more challenging \emph{SITW-DEV} data. Overall, the data-driven components yield now more competitive performance boost over the baseline. Adapting the window function is most effective, with a relative improvement of 9.7\% in EER over the baseline. Concerning DFT, \emph{DFT + loss.} slightly outperforms baseline in both metrics while \emph{DFT + kernel.} is the only variant that does not reach baseline in EER. This finding is in line with \emph{VoxCeleb1-test} results. 
Concerning mel filterbank, all the three systems outperform baseline. Overall, it achieves competitive performance compared with learnable DCT. It reflects its potential on being made adaptable to improve system robustness. 


DET curves for single systems including baseline on \emph{SITW-DEV} have been shown in Fig. \ref{fig:det}. The curves agree with observations from Table \ref{tab:results} in general. Systems with window function adapted produce largest improvement gap with baseline compared with other three, which can be reflected from EER. Considering systems that are less strict on false alarms, we can see that ones like \emph{DCT + kernel.} and \emph{window + loss.} are exceptional and thus can be taken into concern.

\section{Conclusion}
\label{sec:conclusion}

We conduct an initial study on a lightweight learnable MFCC feature extractor as a compromise between end-to-end architectures and hand-crafted feature extractors. 
Our initial results on \emph{SITW-DEV} are promising: the proposed scheme improved upon baseline MFCC extractor. Results for optimized window and mel filterbank are particularly promising. Due to our domain-specific optimization constraints, the learnt representations bear close resemblance to fixed MFCC operations. For interpretability and computational reasons, we restricted the focus on optimization of individual MFCC extractor components; joint optimization of all the four linear components has been left as future work. Similarly, the work can be extended with other deep models such as extended TDNN and ResNet using larger datasets and data augmentation.  



\bibliographystyle{IEEEbib}
\bibliography{strings,refs}

\end{document}